\documentclass[journal,twocolumn]{IEEEtran}
\usepackage{amsfonts}
\usepackage{times}
\usepackage{graphicx}
\usepackage{latexsym}
\usepackage{dsfont}
\usepackage{amssymb}
\usepackage{amsmath}
\usepackage{cite}
\usepackage{verbatim}
\usepackage{subfigure}

\newcommand{\figref}[1]{{Fig.}~\ref{#1}}
\newcommand{\tabref}[1]{{Table}~\ref{#1}}

\def\bb0{{\mathbb{0}}}

\def\bb{{\mathbf{b}}}

\def\bff{{\mathbf{f}}}

\def\bw{{\mathbf{w}}}

\def\b0{{\mathbf{0}}}

\def\bH{{\mathbf{H}}}

\def\cF{\mathcal{F}}

\def\cW{\mathcal{W}}

\def\sf0{{\mathsf{0}}}

\allowdisplaybreaks

\usepackage{epstopdf}
\usepackage{enumerate}
\usepackage{algorithmicx}
\usepackage{algorithm}
\usepackage{amsmath}
\usepackage[noend]{algpseudocode}
\usepackage{float}
\usepackage{color}
\usepackage{makeidx}
\usepackage{rotating}
\usepackage{bbm}
\usepackage{here}
\usepackage{multicol}

\newcommand{\sref}[1]{{Section}~\ref{#1}}

\def \rm {\mathrm}

\begin{document}
\title{Beamforming in Millimeter Wave Systems: Prototyping and Measurement Results}
\author{Cody Scarborough$^\ast$, Kiran Venugopal$^\star$, Ahmed Alkhateeb$^\dagger$,  and  Robert W. Heath, Jr.$^\ddagger$ \\
$^\ast$ University of Michigan, Email: codyscar@umich.edu \\
$^\star$ Qualcomm Corporate R$\&$D,  Email: kiranvg@qti.qualcomm.com\\
$^\dagger$ Arizona State University, Email: alkhateeb@asu.edu \\
$^\ddagger$ The University of Texas at Austin, Email: rheath@utexas.edu\\}
\maketitle

\begin{abstract}
Demonstrating the feasibility of large antenna array beamforming is essential for realizing mmWave communication systems. This is due to the dependency of these systems on the large array beamforming gains to provide sufficient received signal power. In this paper, the design of a proof-of-concept prototype that demonstrates these gains in practice is explained in detail. We develop a mmWave system with digitally controlled analog front-end. The developed prototype uses 60 GHz phased arrays and universal software radio peripheral (USRP) controllers. The software interface of our design is easily reproducible and can be leveraged for future mmWave prototypes and demonstrations. 
\end{abstract}

\section{Introduction} \label{sec:Intro}

To enable millimeter wave (mmWave) systems in practice and to guarantee sufficient received signal power, large antenna arrays need to be employed at both the transmitter and receiver \cite{Rappaport2014,Boccardi2014,Roh2014}. Leveraging the gains of these antennas requires developing beamforming, precoding, and channel estimation algorithms \cite{HeathJr2016}. Further, these antenna array systems have strict hardware constraints that need to be satisfied while developing the beamforming/channel estimation algorithms. Theoretical papers considered various flavors of hybrid MIMO architectures \cite{Mendez-Rial2016}. A main component of these architectures is the digital-controlled analog beamformers. Compared to the variety of mmWave systems in the literature, however, practical implementations in the academia are fewer. This motivates the development of a flexible mmWave design using real hardware in an academic laboratory setting. Such an implementation could then be also used to verify the gains promised by the mmWave systems in theoretical papers.

\subsection{Prior Work}

Some prototypes for the mmWave modem were built by groups at both industry and academia to test the feasibility of beamforming and channel estimation algorithms \cite{Roh2014,Hong2014,Wu2015,Frascolla2015,MiWaveS,Sheldon2009,Brady2013}. An early industrial effort, \cite{Roh2014} developed by Samsung, presented a hybrid beamforming mmWave prototype operating at a carrier frequency of 27.925 GHz. This system had two channels, 4 RF chains per channel and a uniform planar array of 32 antenna elements organized in a subarray structure. This prototype was shown to achieve a peak data rate of 1.056 Gb/s in the laboratory. In the E-band, \cite{Frascolla2015} developed a mmWave backhaul and access prototype to demonstrate the feasibility of mmWave communication and to test the processing algorithms developed under the MiWaveS project \cite{MiWaveS}. In this prototype, the backhaul links employed antennas with 30 dBi gains, and the access links used 17 dBi antenna at the access point (AP) and 5dBi antennas at the user terminals. The antennas at the AP had beamsteering capability of $\pm$45 degrees for both the azimuth and elevation directions. 

To test the feasibility of realizing large antenna arrays at mmWave mobile terminals and its biological implications, \cite{Hong2014} prepared a prototype for a mmWave 5G cellular phone equipped with a pair of 16-element antenna arrays. This study found that the electromagnetic radiation absorbed by a user at 28 GHz is more localized compared to that at 1.9 GHz. The skin penetration depth, however, at 28 GHz is much less -— around 3 mm compared to 45 mm at 1.9 GHz. This implies that most of the absorbed energy is limited to the epidermis at mmWave communications. The biological impact of mmWave radiation has also been further studied in \cite{Wu2015}.

While the industry-developed prototypes in \cite{Roh2014,Hong2014,Wu2015,Frascolla2015,MiWaveS} demonstrated the feasibility of mmWave beamforming in several setups, it is important to develop academic mmWave prototypes to evaluate the different beamforming and channel estimation strategies and provide further insights into the performance of mmWave systems. In \cite{Sheldon2009}, a 60 GHz mmWave prototype was presented in which four LOS spatial multiplexing MIMO links were established to achieve 100 Mbps data rate. In this prototype, both the transmitter and receiver employed four horn antennas, each has 20dBi gain. In \cite{Brady2013}, a beamspace MIMO prototype was built using 40cm x 40cm discrete lens arrays at both the transmitter and receiver. The prototype operates at 10 GHz, and with a separation distance of 2.7m between the communications. This system leveraged spatial multiplexing over LOS links to achieve 10 bps/Hz spectral efficiency at an operating SNR of 32 dB. In \cite{Sheldon2009} and \cite{Brady2013}, the developed prototypes adopted horn antennas and lens arrays. Since phased arrays are expected to be widely used in practical mmWave beamforming systems, it is critical to develop mmWave prototypes based on phased arrays, which is what we consider in this work.

\subsection{Contribution} 
In this paper, we demonstrate how commercially available hardware modules can be integrated and operated to test mmWave MIMO systems, including proposals made in several papers in the literature. We believe that this could pave the way for introducing the study and experimentation of practical mmWave systems into academic curriculum.
\begin{itemize}
	\item We describe how Universal Software Radio Peripherals (USRPs), can be leveraged and attached to mmWave phased arrays, thereby establishing a mmWave communication link. 
	\item We detail all the hardware requirements essential for implementing a mmWave hybrid system and how the mmWave radio can be controlled with a user friendly computer interface. 
	\item We explain how hybrid codebooks can be tested in our proposed prototype using a completely software controlled setup to guarantee sufficient design flexibility. This also enables our prototype to be used for practically validating several precoder/combiner designs and channel estimation techniques available in the literature. 
\end{itemize}

\section{Experimental Setup}
\label{sec:exp_setup}
In this section, we present the hardware components and the setup used to establish the mmWave link, and the completely software-controlled end-to-end system.
\subsection{Hardware Requirements}
\begin{figure}
	\centering
	\subfigure[center][{Phased Array}]{
	    \includegraphics[height=160pt, width=.46\columnwidth]{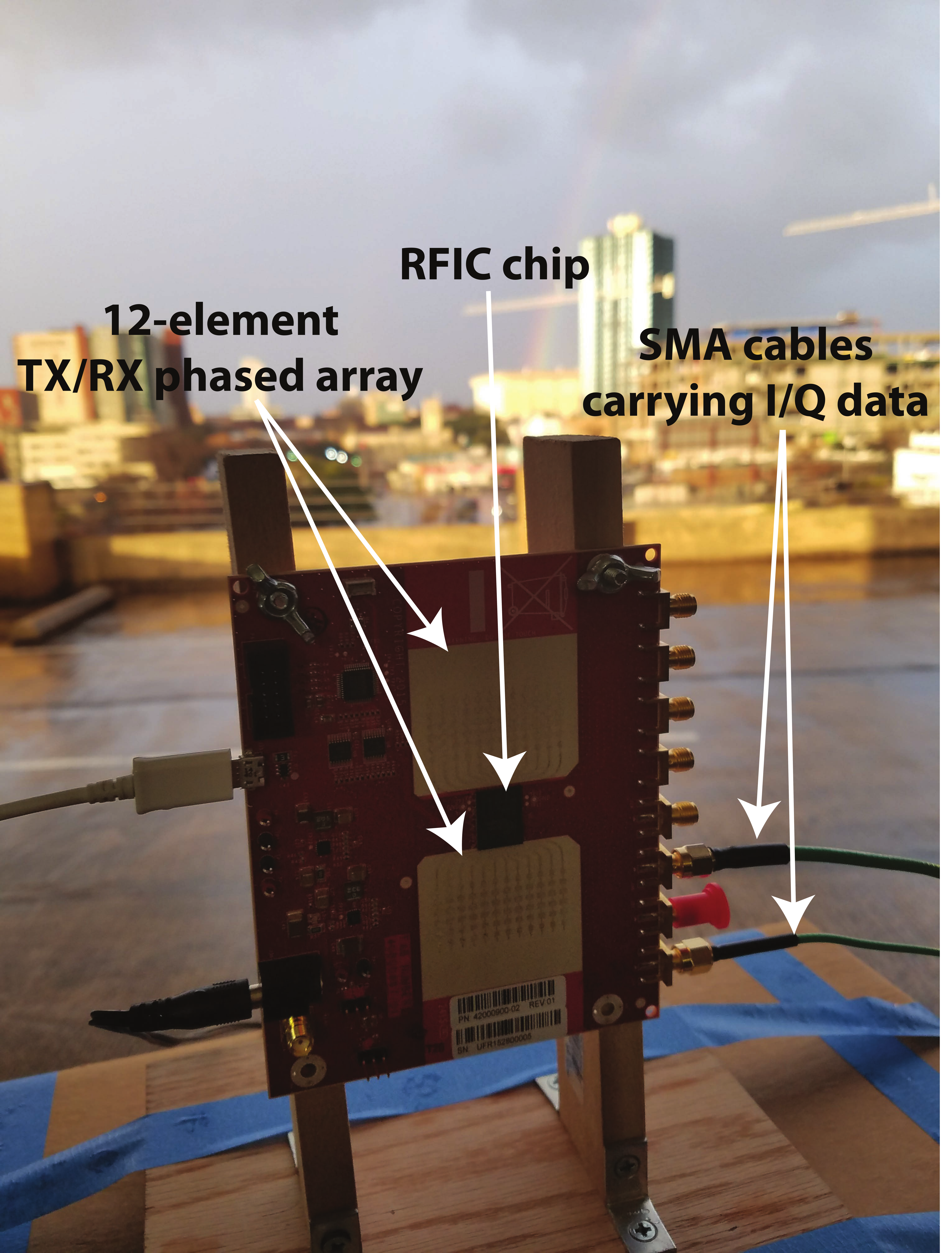}
		\label{fig:PA}}
	\subfigure[center][{Setup}]{
		\includegraphics[height=160pt,width=.46\columnwidth]{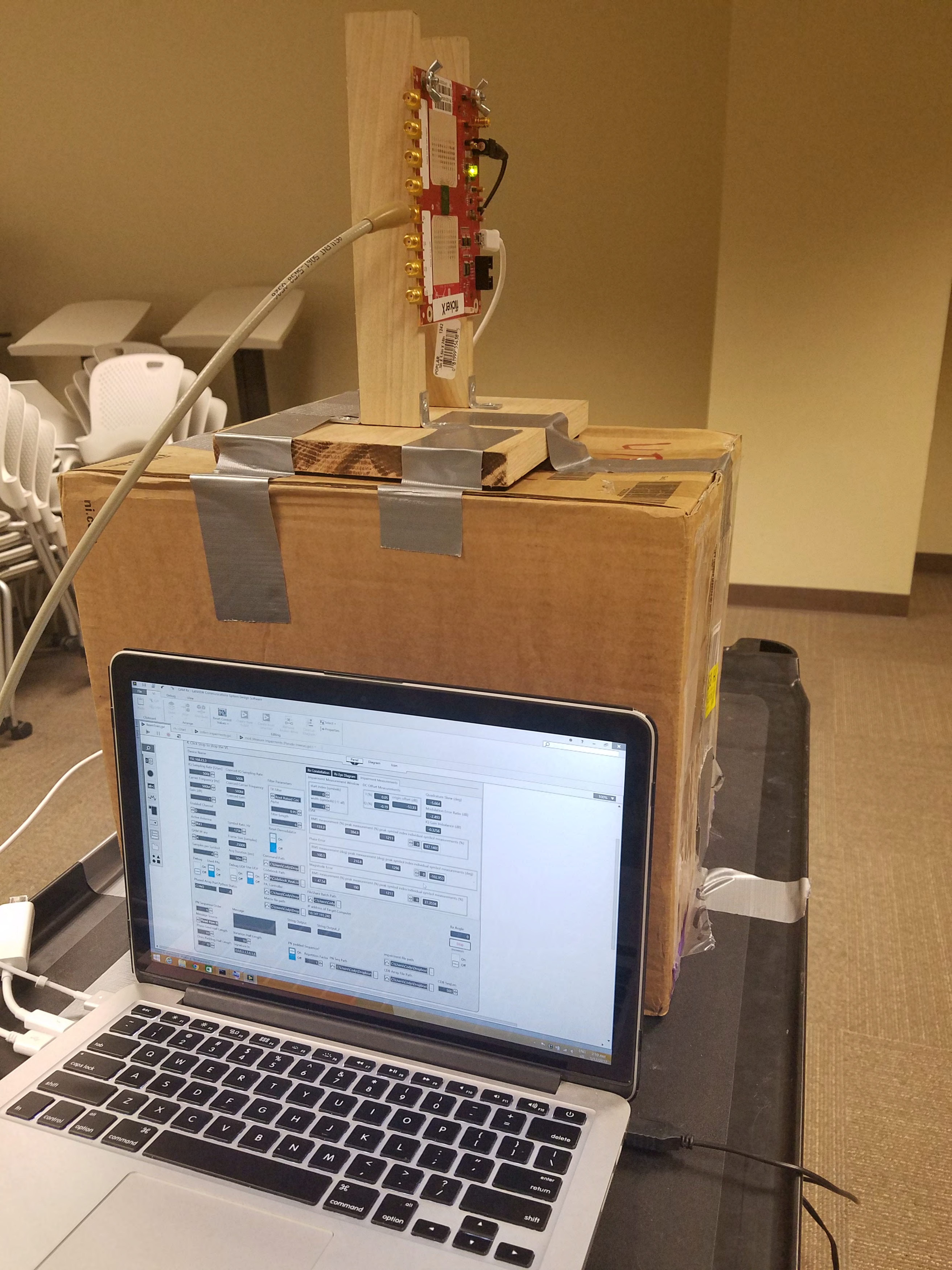}
		\label{fig:Setup1}}
	\caption{Figure showing the phased array donated by Google Inc. used for the prototyping and the proposed setup that can be controlled from a personal computer user interface.}
	\label{Fig:PAandSetup}
\end{figure}

The set of equipment used for the mmWave prototype and channel measurements is summarized in \tabref{tab:hardware_req}. The resulting setup is provided in \figref{Fig:PAandSetup}.
\begin{table}[h]
\caption{Hardware Used}
    \centering
    \begin{tabular}{|c|c|}
        \hline
        Item & Quantity used\\
        \hline
        NI USRP-RIO 2921 & 2\\
        \hline
        Google Phased Arrays & 2\\
        \hline
        Phased Array Mount & 2\\
        \hline
        SMA Cables & 4\\
        \hline
        USB Cables & 2\\
        \hline
        Mobile Carts & 2\\
        \hline
        Ethernet Cables & 2\\
        \hline
        MacBook PRO & 1\\
        \hline
        Dell Laptop & 1\\
        \hline
        Personal Laptop & 1\\
        \hline
        Extension Cord & 1\\
        \hline
    \end{tabular}
    \label{tab:hardware_req}
\end{table}
Next, we describe the key hardware components in more detail. 
\begin{itemize}
    \item \textbf{USRP-RIO 2921:} One USRP each is used at both the transmitter and the receiver sides to connect the phased array boards and the Laptops, where most of the digital signal processing is done. Note that these USRPs are mainly used  as the Analog-to-Digital/Digital-to-Analog Converters, in addition to serving as the intermediate RF units.
    \item \textbf{60 GHz Phased Array:} The custom made phased arrays donated from  Google Inc. had SiBEAM's chip. These operate at 60 GHz and consists of 12 antennas per array, with half-wavelength antenna spacing. Each of the transmitter and receiver is connected to a single phased array. Extensions to assign more than one array per transmitter/receiver and applying hybrid precoding is interesting for future work. 
    \item \textbf{SMA Cables:} SMA cables were used to connect USRPs at the transmitting and receiving ends to the phased array antennas. These carried the transmit/receive signal between the USRPs and the in-phase port of the phased array.
    \item \textbf{USB Cables:} Connection between the controlling laptop and the corresponding phased array was established using USB cables. Through these cables, the phase inputs to each antenna element of the phased array were provided from `codebook' file stored in the computers. 
    \item \textbf{Ethernet Cables:} Ethernet cables were used to communicate between the USRPs and the computers.
    \item \textbf{Laptops:} The transmitter and receiver were connected to two different laptops. These were also used to calibrate the phased array's transmit power. We used a MacBook to interface with the transmitter and a Dell laptop to interface with the receiver. A third laptop was used to remote login to the transmitter and receiver laptops to start and stop the data transmission/reception in the respective laptops. This remote operation is important to avoid the possible signal blockage due to human movement. 
\end{itemize}

\subsection{Hardware Setup}
For data collection, the phased arrays were mounted onto wooden stands and placed on top of carts. This brought the antenna arrays to a height of 1.6 meters above the room floor. The carts also carried the corresponding controlling laptops,the USRPs and connecting cables. Power calibration was required both for the transmitting phased array and the receiving phased array, and consisted of providing an IQ signal to the phased arrays as 30 MHz and 180 $V_{\text{rms}}$. The phased arrays were designed to transmit binary signals at 1 GHz. While the devices available in the lab were unable to generate IQ signals at that rate, the operation of the mmWave link could still be tested by up-converting the available IQ signal to a higher frequency. To this end, two USRP devices were used to take the 10 MHz IQ signal available from the computer and convert this into a 500 MHz intermediate frequency signal. Ethernet cables were used to allow the computers to communicate with these devices. To avoid human interference within the data collection process, Team Viewer was setup on each control laptop and a personal computer was used to control the system remotely.

\begin{figure*}
   \centering
   \includegraphics[width=\textwidth]{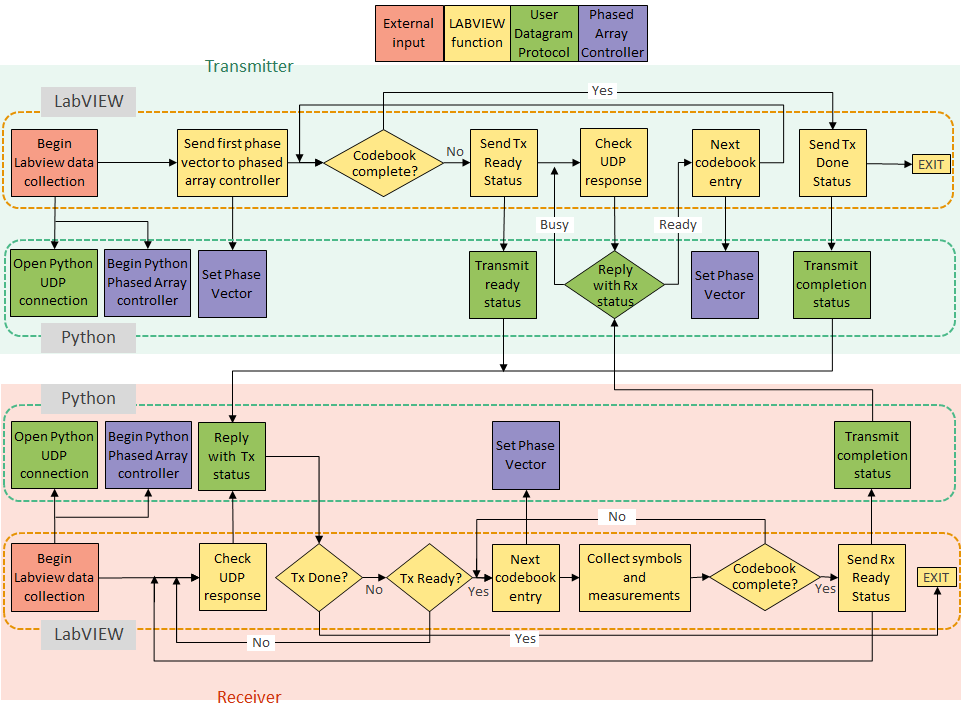}
   \caption{Figure representing the proposed prototype's flow of control between the receiver and transceiver during training via user-defined Python and LabVIEW interfaces.}
  \label{fig:datacollection}
\end{figure*} 

Both Python and LabVIEW were used to control the experimental setup.  A pictorial representation for the flow of control between the Python and LABVIEW parts of the setup is shown in \figref{fig:datacollection}. To begin the data collection process, the LabVIEW virtual instruments (VIs) were run. The transmitter was started first, followed by the receiver. Then user datagram protocol (UDP) connections were initiated at both the transmitter and the receiver using Python. This link served as the connection between the transmitter and receiver used to synchronize the codebook entry selection on both ends. Subsequently, phased array controllers were started at the transceivers. This was followed by sending the first phase vector from the codebook to the transmitter array controller using LabVIEW VI. The corresponding phase vector was also set in the Python part of the control. Once these routines were initialized, the receiving computer set the phase of each element of the phased array to maximize the directivity of the array at the angle specified by the first entry of the receiver codebook. The error vector magnitude (EVM) of the resulting transmission was subsequently obtained and averaged over 50,000 samples (12,500 symbols) at the receiver. This process was repeated for each entry in the receiver codebook until the configurations were exhausted. Following the completion of the receiver codebook, the receiver would send a ready signal to the transmitting computer via UDP. The transmitter would then move on to the next entry in the transmission codebook and reply to the receiver. Once the reply was recognized by the receiver, the data collection cycle repeated. If the transmitter had run out of codebook entries, then a quit command was sent to the receiver and the data collection routine was terminated. In collecting data over various transmitter power configurations, an external loop was created around the collection VI and the process would be repeated over a set of linearly spaced power levels.  To vary the distance between the receiver and transmitter, the cart upon which the transmitter had been placed was manually repositioned to the location of interest, as specified in the experiment. The hardware setup is summarized in \figref{fig:datacollection}.

A central challenge to this work was the validation of the phase array operation. A spectrum analyzer which operated around 60 GHz was unavailable to the authors at the time of testing. This meant that we were unable to  validate the radiation pattern of the antennas via direct measurement. Therefore, we had to rely on the final EVM measurement to determine whether or not the phased array antennas were operating as expected.

\section{System Model}

\subsection{Analog Beamforming}
\label{ssec:AnalogBeamforming}
Using the experimental setup described in \sref{sec:exp_setup}, we use analog beamforming \cite{Hur2013,Wang2009}, to take channel measurements. The digital inputs are given to the phased array to specify the quantized phase angles and thereby control analog beam pointing directions. The analog beam training is performed via exhaustive search as follows. First a given set of phase angles is selected from a codebook to fix the transmit analog beam. With this transmit beam, the receive side beams are swept across the entire codebook. This is then repeated over the entire codebook at the transmitter side to get the measurements corresponding to every possible pairs of transmit and receive analog beam pairs. Mathematically, if $\bH$ denotes the $12 \times 12$ channel matrix between the transmit and receiver phased arrays, and if $\bff, \bw$ represent the transmit and receive beamforming vectors, then the objective of this exhaustive search is to find $\bff^\star$ and $\bw^\star$ that solve 
\begin{align} \label{eq:Opt1}
\left\{\bff^\star, \bw^\star\right\} = & \arg\max_{\substack{ \bff \in \boldsymbol{\cF} \\ \bw \in \boldsymbol{\cW}}} \left|\bw^H \bH \bff\right|, 
\end{align}
where $\boldsymbol{\cF}$ and $\boldsymbol{\cW}$ are the precoding and combining codebooks that gather all the candidate transmit and receive beamforming vectors. It is important to note here that implementing the transmit and receive beamforming vectors using a network of quantized phase shifters is the main reason why these beamforming vectors can take only a finite set of values. These codebooks, though, could also be useful for limited feedback operation \cite{Niu2017,Alkhateeb2016b}. An alternative to exhaustive codebook search is the hierarchical beam training approach, which we save for future work.
 
\subsection{Experimental Environment}
\begin{figure}
   \centering
   \includegraphics[width=1\columnwidth]{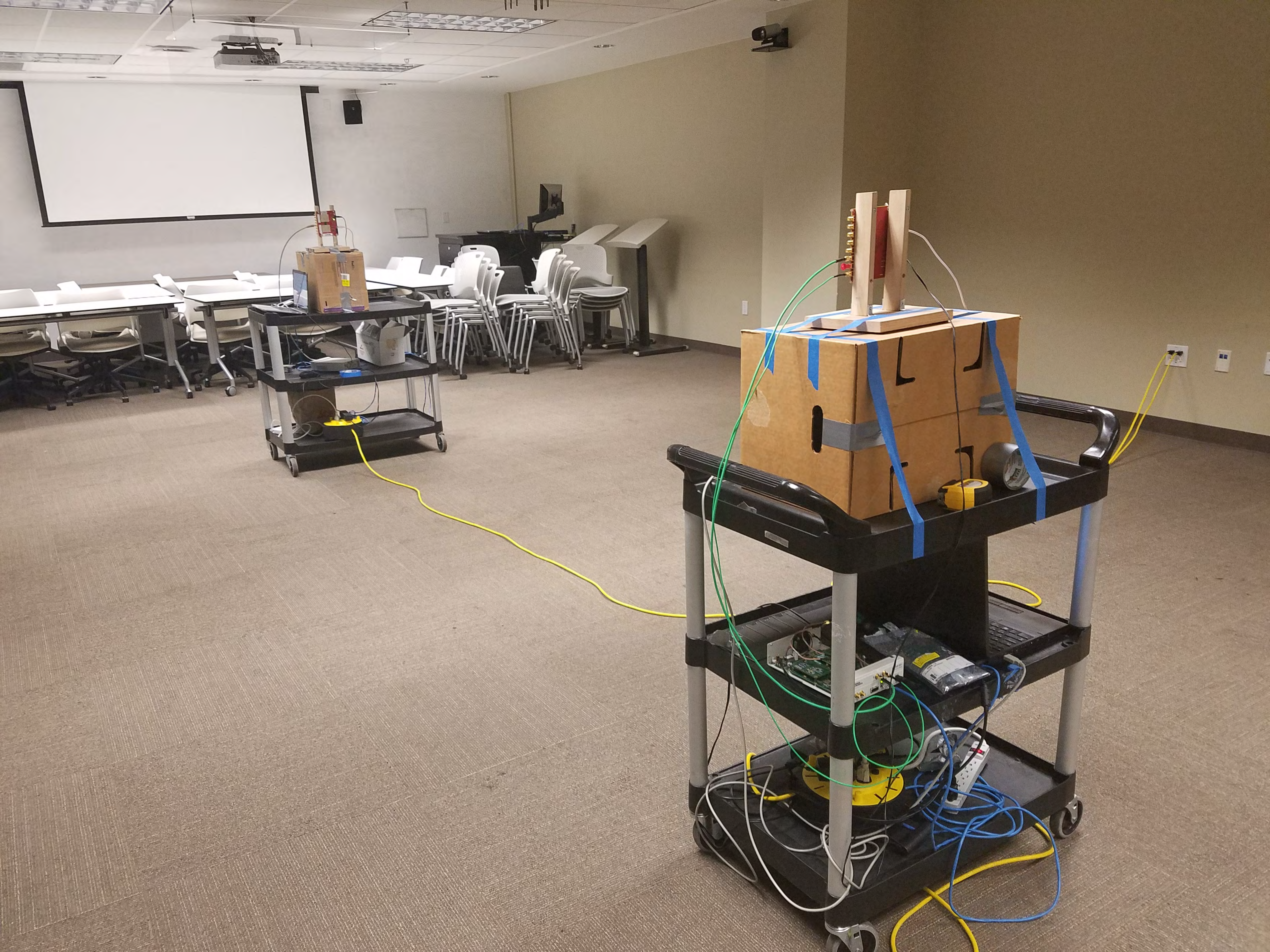}
   \caption{Picture of the indoor environment where the measurements were taken using the proposed mmWave prototype.}
  \label{fig:lab_env}
\end{figure}

The testing environment was a rectangular lecture room, 5 meters wide and 10 meters long. The furniture was moved aside to avoid any potential signal blockage, and thus clearing out the room center. In future experiments, the impact of the positioning and density of the furniture can also be investigated. In our current setup, the receiver antenna was located at the center of the transverse dimension of the room and 2 meters away from the back wall. The floor of the room was carpet, and the foam tiled ceiling was approximately 3 meters above the floor. During the measurement process, the transmitter was placed at the center of the transverse dimension of the wall, and moved along the longitudinal direction to achieve a specific distance separating the transmitting array and receiving array. Apart from human involvement to accurately position the transceivers and to measure the distance between the transmitter and receiver carts, caution was taken to let no mechanical or human movements during the channel measurement process within the lecture room. This was enabled by the software control and the flexible VI design to automate the transmission and reception of data symbols, while digitally controlling the beamforming codebook. The experimental environment with the hardware setup positioned at the center of the lecture room is shown in \figref{fig:lab_env}.

\section{Measurement Results}
In this section, measurement results obtained using our mmWave beamforming prototype are presented. As a preliminary work to show how the mmWave MIMO prototype can be leveraged in practice, we mainly looked at the error vector magnitude (EVM) at the receiver for the various beamformers specified by the codebook and at different link distances between the transmitter and the receiver. This could be easily extended to consider other performance metrics such as block error rates, received signal-to-noise-ratio, achievable spectral efficiency etc.

To evaluate the EVM, we first computed the RMS error vectors' amplitude $E_{\text{error}}$, averaging over multiple experiment runs and data transmission. Then the receiver EVM is computed relative to the reference symbols' RMS amplitude $E_{\text{ref}}$ as 
\begin{eqnarray}
\text{EVM} = 10\log\left(\frac{E_{\text{error}}}{E_{\text{ref}}}\right).
\end{eqnarray}

\begin{figure}
	\centerline{
		\includegraphics[scale=.55]{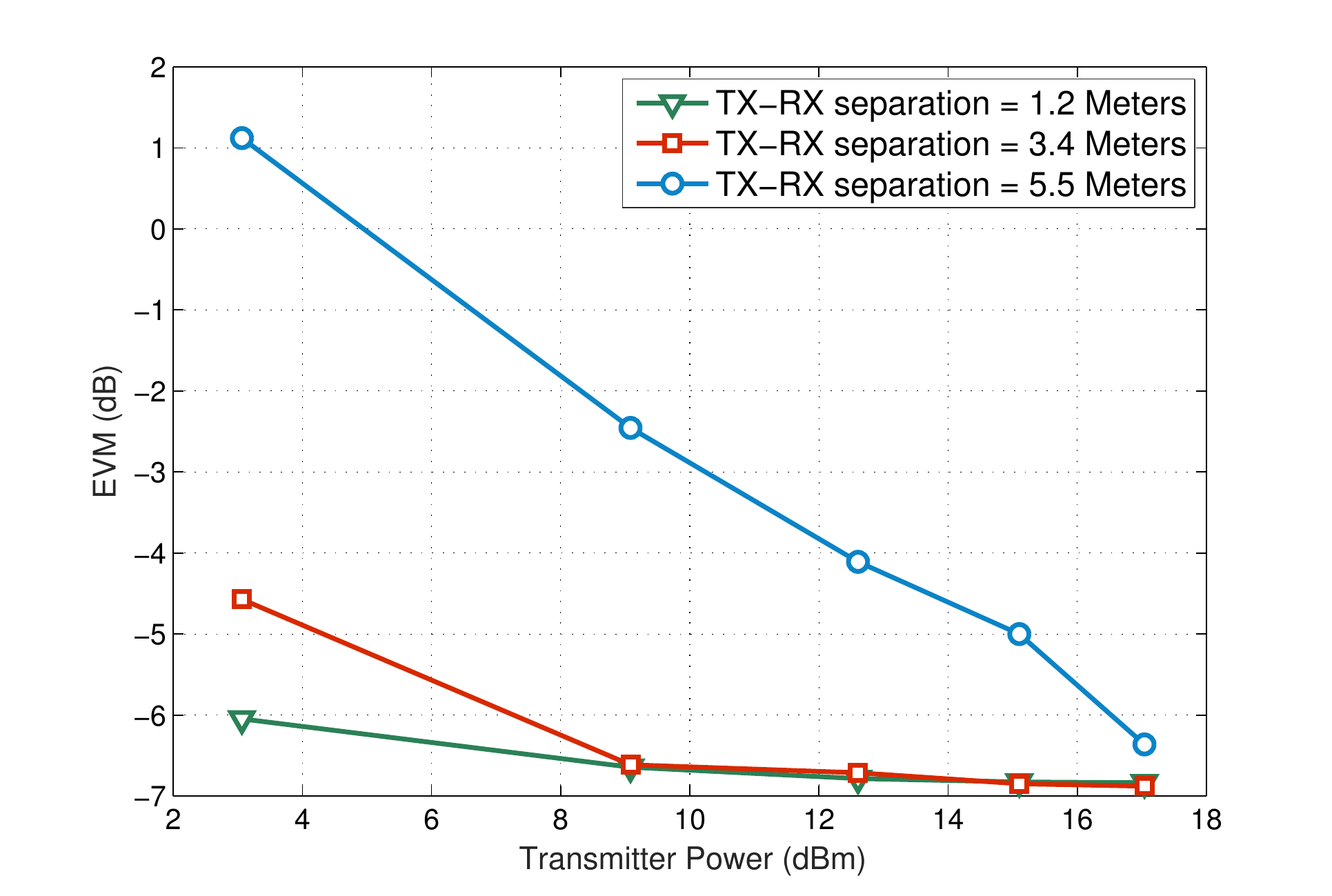}
	}
	\caption{Figure showing the error vector magnitude as a function of transmit power for various link distances.}
	\label{fig:Model}
\end{figure}

\figref{fig:Model} shows the observed EVM as a function of the transmit power in dBm for various link distances between the transmitter and the receiver. Importantly, the plotted EVM values correspond to the best beamforming codeword used at the transmitter and the receiver, which is determined from the measured EVM for every possible combination of precoder and combiner weights specified by the hybrid codebook. In short, in this paper, we used EVM as the metric to evaluate the best transmit and receive beamforming vectors as described in \sref{ssec:AnalogBeamforming}. As expected, in \figref{fig:Model}, a degradation in performance is observed in the EVM as the separation between the transmitter and the receiver is increased. The distance dependent pathloss of the wireless link accounts for this observation. Further, the EVM is also seen to improve as we increase the transmit power, confirming the usual trend, since increasing the transmit power effectively increases the RMS amplitude of the reference symbols.

\section{Acknowledgment} 
This work was done while the first three authors were at The University of Texas at Austin. This material is based upon work supported in part by the National Science Foundation under Grant Nos. ECCS-1711702 and CNS-1731658, and the U.S. Department of Transportation through the Data-Supported Transportation Operations and Planning (D-STOP) Tier 1 University Transportation Center.

\section{Conclusion} \label{sec:Conc}
In this paper, we presented details of how a mmWave beamforming prototype based on USRPs and 60 GHz phased arrays can be implemented in practice. Importantly, this prototype can be setup in the wireless labs of the  academia to further study mmWave wireless systems. The system design and the user interface are flexible and can be directly used to evaluate the efficiency of the RF codebook-based beamforming techniques that assume digitally controlled phased arrays. As a future extension, it is interesting to develop a phased-array based prototype with multiple RF chains that is capable of evaluating hybrid beamforming and channel estimation strategies. Multiple such setups can also be used to study multi-user performance of hybrid mmWave systems.

\bibliographystyle{IEEEtran}

\end{document}